\documentclass{PoS}
\usepackage{graphicx,pstricks,rotating,amsmath}
\usepackage{slashed,epstopdf}

\title{Photon polarization tensor on deformed spacetime: A four-photon-tadpole contribution\footnote{See also the published article~\cite{Horvat:2015aca} of these results.}}

\ShortTitle{Photon Self-Interaction in a Deformed $\rm U(1)$ Gauge Theory}

\author{Raul Horvat\\
        Institute Rudjer Bo\v{s}kovi\'{c}, Bijeni\v{c}ka 54 10000 Zagreb, Croatia\\
        E-mail: \email{raul.horvat@irb.hr}}

\author{Josip Trampeti\'{c}\thanks{On leave of absence from Institute Rudjer Bo\v{s}kovi\'{c}, Zagreb, Croatia}\\
        Max-Planck-Institut f\"ur Physik, (Werner-Heisenberg-Institut), F\"ohringer Ring 6, D-80805 M\"unchen, Germany\\
        E-mail: \email{trampeti@mppmu.mpg.de,josipt@rex.irb.hr
}}

\author{\speaker{Jiangyang You}
\\
        Institute Rudjer Bo\v{s}kovi\'{c}, Division of Theoretical Physics, Bijeni\v{c}ka 54 10000 Zagreb, Croatia\\
        E-mail: \email{youjiangyang@gmail.com}}

\newcommand{\tr}{\hbox{tr}}

\abstract{In this contribution to the Proceedings of the Corfu Summer Institute 2015 we present our study on photon self-interaction in the deformed U(1) gauge field theory defined via Seiberg-Witten map. We derive the $\theta$-exact expressions for the four photon self-coupling in this theory with a large number of gauge symmetry inspired freedom parameters included. The physical effect of four photon self-coupling is accessed by its contribution to photon polarization tensor via 1-loop four-photon-tadpole-diagram. The four-photon tadpole diagram consists the same tensorial structures as the photon self-interacting bubble diagram we studied before, with however only quadratically IR divergent coefficients. We show that there exists a unique combination of gauge symmetry inspired freedoms which induces {\em full} quadratic IR divergence cancellation for {\em arbitrary} noncommutative parameter $\theta^{ij}$ once the photon self-interacting bubble and tadpole diagrams are summed over.}

\FullConference{Proceedings of the Corfu Summer Institute 2015 "School and Workshops on Elementary Particle Physics and Gravity, 1-27 September 2015, Corfu, Greece}

\begin{document}

\section{Introduction}

After two decades of development, many open questions still remain in the field of the perturbative quantization of the noncommutative (NC) field theories on Moyal space~\cite{Blaschke:2016gxl}. One of most renowned issues is the quadratic infrared (IR) divergence in the one-loop 1-PI two point function of the gauge fields~\cite{Hayakawa:1999yt,Hayakawa:1999zf,Matusis:2000jf}. Actually its existence in a very large and important subcategory of the deformed gauge theories on Moyal space, namely those defined via Seiberg-Witten (SW) map~\cite{Seiberg:1999vs,Jurco:2000fb,Madore:2000en,Jurco:2000ja,Mehen:2000vs,Jurco:2001my,Jurco:2001rq,Barnich:2002pb,Barnich:2003wq,Banerjee:2003vc,Banerjee:2004rs,Martin:2012aw,Martin:2015nna,Aschieri:2012in,Aschieri:2014xka,AschieriCorfu}, was confirmed not long ago~\cite{Schupp:2008fs,Horvat:2013rga}. Unlike the quadratic UV/IR mixing in the scalar field theory on Moyal space~\cite{Minwalla:1999px}, this quadratic IR divergence does not have a quadratic UV divergent counterpart, and appears to be difficult to control~\cite{Blaschke:2008yj,Blaschke:2009aw,Blaschke:2009gm,Blaschke:2009zi,Blaschke:2016gxl}.

In~\cite{Horvat:2013rga} the one loop photon-bubble-diagram contributions to the photon polarization tensor $\mathcal B^{\mu\nu}(p)$ was studied in a $\theta$-exact SW map expanded U(1) gauge theory. Five tensor structures were found after the loop integrals are evaluated using generalized dimensional regularization procedure:
\begin{equation}
\begin{split}
\mathcal B^{\mu\nu}(p)&=\frac{e^2}{(4\pi)^2}\bigg\{\Big[g^{\mu\nu}p^2-p^\mu p^\nu\Big]B_1(p)
+(\theta p)^\mu (\theta p)^\nu B_2(p)
\\&+\Big[g^{\mu\nu}(\theta p)^2-(\theta\theta)^{\mu\nu}p^2
+ p^{\{\mu}(\theta\theta p)^{\nu\}}\Big]B_3(p)
\\&+\Big[(\theta\theta)^{\mu\nu}(\theta p)^2+(\theta\theta p)^\mu(\theta\theta p)^\nu\Big]B_4(p) + (\theta p)^{\{\mu} (\theta\theta\theta p)^{\nu\}} B_5(p)\bigg\},
\end{split}
\label{4.1}
\end{equation}
with the following divergent parts of the $B_i$-coefficients
\begin{gather}
\begin{split}
B_1(p)&\sim+\bigg(\frac{1}{3}\big(1-3\kappa\big)^2 +\frac{1}{3}\big(1+2\kappa\big)^2\; \frac{p^2(\tr\theta\theta)}{(\theta p)^2}
+\frac{2}{3}\big(1+4\kappa+\kappa^2\big)\; \frac{p^2(\theta\theta p)^2}{(\theta p)^4}\bigg)
\\&\hspace{5.5mm}\cdot\left[\frac{2}{\epsilon} + \ln(\mu^2(\theta p)^2)\right]
\hspace{3mm}-\frac{8}{3}\big(1-\kappa\big)^2\;\frac{1}{(\theta p)^6}\bigg((\tr\theta\theta)(\theta p)^2+4(\theta\theta p)^2\bigg)\,,
\\
B_2(p)&\sim+\bigg(\frac{4}{3}\big(1-\kappa\big)^2\; \frac{p^4(\theta\theta p)^2}{(\theta p)^6}+\frac{1}{3}\big(1-2\kappa-5\kappa^2\big)\frac{p^4(\tr\theta\theta)}{(\theta p)^4}+\frac{1}{3}\big(25
\\&\hspace{4.5mm}-86\kappa+73\kappa^2\big)\frac{p^2}{(\theta p)^2}\bigg)\left[\frac{2}{\epsilon} + \ln(\mu^2(\theta p)^2)\right]
-\frac{8}{3}\big(1-3\kappa\big)\big(3-\kappa\big)\frac{1}{(\theta p)^4}
\\&\hspace{4.5mm}+\frac{16}{3}(1-\kappa\big)^2\frac{1}{(\theta p)^8}\bigg((\tr\theta\theta)(\theta p)^2+6(\theta\theta p)^2\bigg),
\\
B_3(p)&\sim-\frac{1}{6}\big(1-2\kappa-11\kappa^2\big)\frac{p^2}{(\theta p)^2}
\left[\frac{2}{\epsilon} + \ln(\mu^2(\theta p)^2)\right]
-\frac{4}{3(\theta p)^4}\big(1-10\kappa+17\kappa^2\big),
\\
B_4(p)&\sim-\big(1+\kappa\big)^2\frac{p^4}{(\theta p)^4}\left[\frac{2}{\epsilon} + \ln(\mu^2(\theta p)^2)\right]-\frac{16p^2}{3(\theta p)^6}\big(1-6\kappa+7\kappa^2\big),
\\
B_5(p)&\sim+\frac{2}{3}\big(1+\kappa+4\kappa^2\big)\frac{p^4}{(\theta p)^4}
\left[\frac{2}{\epsilon} + \ln(\mu^2(\theta p)^2)\right]+\frac{32p^2}{3(\theta p)^6}\big(1-\kappa\big)\big(1-2\kappa\big).
\end{split}
\label{4.2}
\end{gather}
One can easily observe the coexistence of both UV and quadratic IR divergences in the $B_i$s. The parameter $\kappa$ here presents the the first ($e^2$) order SW map amibiguity/freedom of the gauge field strenght. This concept was originally introduced via a $\theta$-iterative construction~\cite{Bichl:2001cq} as a countermeasure to the infinite series of UV divergences there. Our formulism here follows a $\theta$-exact substitute suggested in~\cite{Trampetic:2014dea}. It is further shown in~\cite{Horvat:2013rga} that $\kappa$ can provide full control over both UV and IR divergences in the bubble diagram only when a special full rank value of the Moyal $\theta^{ij}$ tensor is selected.

Despite significant progresses made in~\cite{Horvat:2013rga} the result is incomplete because there also exists four photon self-coupling induced by the second order $\theta$-exact Seiberg-Witten map expansion, which is, historically, largely an untouched subject due to mathematical sophistication. Following the recent results on higher order $\theta$-exact Seiberg-Witten map~\cite{Martin:2012aw,Trampetic:2015zma,Martin:2015nna}, we present here the full $\theta$-exact four photon self-interaction in the SW mapped deformed pure U(1) gauge theory on Moyal space and the evaluation of the resulting four-photon tadpole diagram, thus complete the one loop corrections to the photon polarization tensor.

Our explicit computation shows that photon tadpole diagram produces the same five tensor structures as the photon bubble diagrams. The tadpole integrals are, however, purely quadratic IR divergent. Therefore they could be the origin of all quadratic IR divergences in noncommutative gauge theories on Moyal space, while the absence of a UV counterpart is then explained by the vanishing of the commutative tadpole integrals in the dimensional regularization~\cite{Leibbrandt:1975dj}. We have also included a series of new gauge symmetry inspired freedom parameters $\kappa_i$s as the second order extension of parameter $\kappa$ in our model definition, which are shown to offer full control over only the quadratic IR divergences for arbitrary values of Moyal $\theta^{ij}$ tensor together with the first order parameter $\kappa$.

The article is structured as follows: The four photon self-interaction is defined in the Section 2. Section 3 handles the corresponding Feynman rules and the resulting tadpole diagram. The full one loop quadratic IR divergences in the photon polarization tensor is presented in the Section 4, then follow the discussion and conclusions. Note that in this article the capital letters denote noncommutative objects, while the small letters denote the commutative ones.

\section{Model definition}

We consider the formal $\rm U_{\star}(1)$ NC gauge theory action
\begin{equation}
S=-\frac{1}{4e^2}\int F_{\mu\nu}\left(e\cdot a_\mu,\theta^{\mu\nu}\right)
\star F^{\mu\nu}\left(e\cdot a_\mu,\theta^{\mu\nu}\right),
\label{2.00}
\end{equation}
where the formal NC gauge field strength
$F_{\mu\nu}\left(e\cdot a_\mu,\theta^{\mu\nu}\right)$
is regarded as a composite operator built-up using the commutative gauge field operator $a_\mu$ and the NC parameter $\theta^{\mu\nu}$ via the SW map procedure. The commutative coupling constant $e$ is attached to the commutative gauge field operator $a_\mu$ due to the charge quantization issue~\cite{Horvat:2011qn}. As a bonus feature it also serves as the ordering parameter for the $\theta$-exact SW map expansion, i.e.
\begin{equation}
F_{\mu\nu}\left(e\cdot a_\mu,\theta^{\mu\nu}\right)=e f_{\mu\nu}+F^{e^2}_{\mu\nu}+F^{e^3}_{\mu\nu}+\mathcal O\left(e^4\right)
\label{2.0}.
\end{equation}
The $e^2$ order gauge field strength Seiberg-Witten map $F^{e^2}_{\mu\nu}$ expansion is fairly universal
\begin{equation}
F^{e^2}_{\mu\nu}=e^2\theta^{ij}
\Big(\kappa f_{\mu i}\star_{2}f_{\nu j}-a_i\star_2\partial_j
f_{\mu\nu}\Big).
\label{2.1}
\end{equation}
The structure of the $\theta$-exact SW map of a $\rm U(1)$ gauge theory is summarized in \cite{Trampetic:2015zma}, where two distinct gauge field SW maps were found and analyzed up to the $e^3\sim a_\mu^4$ order.

Expanding \eqref{2.00} to order $a_\mu^4$ gives the following general form for the photon self-interaction
\begin{equation}
S^{e^2}=-\frac{1}{4e^2}\int\,F^{e^2}_{\mu\nu}F^{ e^2\mu\nu}+2ef^{\mu\nu}F_{\mu\nu}^{e^3},
\label{2.2}
\end{equation}
where the following distinct solutions for the $e^3$ order gauge field strength have been found and given explicitly in \cite{Trampetic:2015zma}. The first one is resolved from Seiberg-Witten differential equation:
\begin{equation}
\begin{split}
F^{e^3}_{\mu\nu}&(x)_{\kappa,\kappa_1,\kappa_2}=
\frac{e^3}{2}\theta^{ij}\theta^{kl}\bigg[\kappa_1\left(\left[f_{\mu k}f_{\nu i} f_{l j}\right]_{\star_{3'}}+\left[f_{\nu l}f_{\mu i}f_{kj}\right]_{\star_{3'}}\right)-\kappa a_i\star_2\partial_j\left(f_{\mu k}\star_2 f_{\nu l}\right)
\\&-\kappa_2\left(\left[f_{\nu l}a_i\partial_j f_{\mu k}\right]_{\star_{3'}}
+\left[f_{\mu k}a_i\partial_j f_{\nu l}\right]_{\star_{3'}}+\left[a_k\partial_l\left(f_{\mu i}f_{\nu j}\right)\right]_{\star_{3'}}-2a_i\star_2\partial_j\left(f_{\mu k}\star_2 f_{\nu l}\right)\right)
\\&+\left[a_i\partial_j a_k \partial_l f_{\mu\nu}\right]_{\star_{3'}}
+\left[\partial_l f_{\mu\nu}a_i\partial_j a_k\right]_{\star_{3'}}+\left[a_k a_i \partial_l\partial_j f_{\mu\nu}\right]_{\star_{3'}}
\\&-\frac{1}{2}\Big(\left[a_i\partial_k a_j\partial_l f_{\mu\nu}\right]_{\star_{3'}}
+\left[\partial_l f_{\mu\nu}a_i\partial_k a_j\right]_{\star_{3'}}\Big)\bigg]\,.
\label{2.3}
\end{split}
\end{equation}

The generalized star products are defined via the following modified Fourier transformations
\begin{gather}
\begin{split}
(f\star_2 g)(x)&=\int\,e^{-i(p+q)x}\tilde f(p)\tilde g(q)f_{\star_2}\left(p,q\right),
\\
[fgh]_{\star_3}(x)&=\int\,e^{-i(p+q+k)x}\tilde f(p)\tilde g(q)\tilde h(k)f_{\star_3}\left(p,q,k\right),
\\
[fgh]_{\star_{3'}}(x)&=\int\,e^{-i(p+q+k)x}\tilde f(p)\tilde g(q)\tilde h(k)f_{\star_{3'}}\left(p,q,k\right).
\label{2.7}
\end{split}
\end{gather}
The definitions for the momentum dependent
functions $f_{\star_2}\left(p,q\right)$, $f_{\star_3}\left(p,q,k\right)$
and $f_{\star_{3'}}\left(p,q,k\right)$ are given in the Appendix A.

In the above solutions we have included the following freedom parameters: $\kappa$ in $F^{e^2}_{\mu\nu}$, while for $F^{e^3}_{\mu\nu}$ we have $(\kappa,\kappa_{1,2})$
 From those field strengths we have found the following two actions at the $a_\mu^4$ order,
\begin{equation}
\begin{split}
S^{e^2}=&-\frac{e^2}{4}\theta^{ij}\theta^{kl}\int\,\kappa^2(f_{\mu i}\star_2 f_{\nu j})(f^\mu_{\;\,\; k}\star_2 f^\nu_{\;\,\; l})-\kappa(f_{ij}\star_2 f_{\mu\nu})(f^\mu_{\;\,\; k}\star_2 f^\nu_{\;\,\; l})
\\&+2\kappa_1f^{\mu\nu}[f_{\mu i}f_{\nu k}f_{jl}]_{\star_{3'}}
\\&+2\kappa_2 f^{\mu\nu}\left(a_i\star_2\partial_j(f_{\mu k}\star_2 f_{\nu l})-[f_{\mu k} a_i\partial_j f_{\nu l}]_{\star_{3'}}-[a_i f_{\mu k}\partial_j f_{\nu l}]_{\star_{3'}})\right)
\\&+(a_i\star_2\partial_j f_{\mu\nu})(a_k\star_2\partial_l f^{\mu\nu})
\\&+\frac{1}{2}f^{\mu\nu}\left(2[a_i\partial_j a_k \partial_l f_{\mu\nu}]_{\star_{3'}}+2[\partial_l f_{\mu\nu}a_i \partial_j a_k]_{\star_{3'}}+2[a_i a_k \partial_j\partial_l f_{\mu\nu}]_{\star_{3'}}
\right.\\&\left.-[a_i\partial_k a_j\partial_l f_{\mu\nu}]_{\star_{3'}}-[\partial_l f_{\mu\nu}a_i\partial_k a_j]_{\star_{3'}}\right).
\end{split}
\label{2.8}
\end{equation}
It is possible to express the action \eqref{2.3} fully in terms of the commutative field
strength $f_{\mu\nu}$ by applying a large number of integrations-by-part on the
relevant terms. The outcomes 
are given below:
\begin{equation}
\begin{split}
S^{e^2}=&-\frac{e^2}{4}\theta^{ij}\theta^{kl}\int\,\kappa^2(f_{\mu i}\star_2 f_{\nu j})(f^\mu_{\;\,\; k}\star_2 f^\nu_{\;\,\; l})-\kappa(f_{ij}\star_2 f_{\mu\nu})(f^\mu_{\;\,\; k}\star_2 f^\nu_{\;\,\; l})
\\&+2\kappa_1 f^{\mu\nu}\left(a_i\star_2\partial_j(f_{\mu k}\star_2 f_{\nu l})-[f_{\mu k} a_i\partial_j f_{\nu l}]_{\star_{3'}}-[a_i f_{\mu k}\partial_j f_{\nu l}]_{\star_{3'}})\right)
\\&+2\kappa_2f^{\mu\nu}[f_{\mu i}f_{\nu k}f_{jl}]_{\star_{3'}}
\\&-\frac{1}{4}f^{\mu\nu}\left[f_{\mu\nu}f_{ik}f_{jl}\right]_{\star_{3'}}+\frac{1}{8}\left(f^{\mu\nu}\star_2 f_{ij}\right)\left(f_{kl}\star_2 f_{\mu\nu}\right)
\\&+\frac{1}{2}\theta^{pq}f_{\mu\nu}\left[\partial_i f_{jk} f_{lp}\partial_q f_{\mu\nu}\right]_{\mathcal M_{\rm (I)}}.
\end{split}
\label{2.10}
\end{equation}

The products $\mathcal M_{\rm (I)}$, and $\mathcal M_{\rm (II)}$ needed later, are defined via the momentum structures $f_{\rm (I,II)}$ given in Appendix A:
\begin{equation}
[fgh]_{\mathcal M_{\rm (I,II)}}(x)=\int\,e^{-i(p+q+k)x}\tilde f(p)\tilde g(q)\tilde h(k)f_{\rm (I,II)}\left(p,q,k\right).
\label{2.12}
\end{equation}

Note that terms of order $\theta^2$, $\theta^{ij}\theta^{kl}f_{ik}f_{jl}f_{\mu\nu}$ and
$\theta^{ij}\theta^{kl}f_{ij}f_{kl}f_{\mu\nu}$, can be generated via the $\theta$-iterative
procedure ~\cite{Bichl:2001cq}. We thus introduce in first model two additional freedom
parameters $(\kappa_3, \kappa_4)$ (and $(\kappa'_3, \kappa'_4)$ in model (II)), as the $\theta$-exact completion of these two freedoms. In this way we  produce the following final forms for the $a_{\mu}^4$-order action:
\begin{equation}
\begin{split}
S^{e^2}_{\kappa,\kappa_1,\kappa_2,\kappa_3,\kappa_4}=&-\frac{e^2}{4}\theta^{ij}\theta^{kl}\int\,\kappa^2(f_{\mu i}\star_2 f_{\nu j})(f^\mu_{\;\,\; k}\star_2 f^\nu_{\;\,\; l})-\kappa(f_{ij}\star_2 f_{\mu\nu})(f^\mu_{\;\,\; k}\star_2 f^\nu_{\;\,\; l})
\\&+2\kappa_1 f^{\mu\nu}\left(a_i\star_2\partial_j(f_{\mu k}\star_2 f_{\nu l})-[f_{\mu k} a_i\partial_j f_{\nu l}]_{\star_{3'}}-[a_i f_{\mu k}\partial_j f_{\nu l}]_{\star_{3'}})\right)
\\&+2\kappa_2f^{\mu\nu}[f_{\mu i}f_{\nu k}f_{jl}]_{\star_{3'}}
\\&-\frac{\kappa_3}{4}f^{\mu\nu}\left[f_{\mu\nu}f_{ik}f_{jl}\right]_{\star_{3'}}+\frac{\kappa_4}{8}\left(f^{\mu\nu}\star_2 f_{ij}\right)\left(f_{kl}\star_2 f_{\mu\nu}\right)
\\&+\frac{1}{2}\theta^{pq}f_{\mu\nu}\left[\partial_i f_{jk} f_{lp}\partial_q f_{\mu\nu}\right]_{\mathcal M_{\rm (I)}}.
\end{split}
\label{2.15}
\end{equation}

\section{Vertex-diagram-loop integral}

\begin{figure}
\begin{center}
\includegraphics[width=4cm,height=2cm]{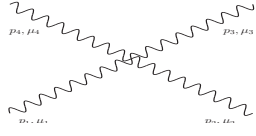}
\end{center}
\caption{Four-photon field vertex $\Gamma^{\mu_1\mu_2\mu_3\mu_4}(p_1,p_2,p_3,p_4)$ with all incomming momenta.}
\label{fig:vertex}
\end{figure}

From the action \eqref{2.15} we read out the corresponding
four-photon interactions in the momentum space, with all four
momenta $p_i$ in Fig.\ref{fig:vertex} being the incoming ones
\begin{equation}
\begin{split}
\Gamma^{\mu_1\mu_2\mu_3\mu_4}\left(p_1,p_2,p_3,p_4\right)=&-i\frac{e^2}{4}\Big(\kappa^2\Gamma_{\rm A}^{\mu_1\mu_2\mu_3\mu_4}\left(p_1,p_2,p_3,p_4\right)
\\&+\kappa\Gamma_{\rm B}^{\mu_1\mu_2\mu_3\mu_4}\left(p_1,p_2,p_3,p_4\right)
+\kappa_1\Gamma_1^{\mu_1\mu_2\mu_3\mu_4}\left(p_1,p_2,p_3,p_4\right)\\&+\kappa_2\Gamma_2^{\mu_1\mu_2\mu_3\mu_4}\left(p_1,p_2,p_3,p_4\right)
+\kappa_3\Gamma_3^{\mu_1\mu_2\mu_3\mu_4}\left(p_1,p_2,p_3,p_4\right)
\\&+\kappa_4\Gamma_4^{\mu_1\mu_2\mu_3\mu_4}\left(p_1,p_2,p_3,p_4\right)+\Gamma_5^{\mu_1\mu_2\mu_3\mu_4}\left(p_1,p_2,p_3,p_4\right)\Big)
\\&+{\rm all\; S_4\; permutations\; over}\; \{p_i\}{\rm \; and}\; \{\mu_i\}{\rm \; simutaneously}.
\end{split}
\label{3.1}
\end{equation}

The definition of $\Gamma_{\rm A}$, $\Gamma_{\rm B}$, $\Gamma_i$ and $\Gamma'_i$ 's are given in~\cite{Horvat:2015aca}.

\subsection{Photon two-point function: Four-photon-tadpole diagram}

The photon-loop computation involves a single 4-photon-tadpole-loop integral contribution to the photon polarization tensor in $D$ dimensions
\begin{figure}
\begin{center}
\includegraphics[width=5cm,height=3cm]{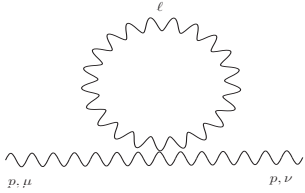}
\end{center}
\caption{Four-photon-tadpole contribution to the photon two-point function $T^{\mu\nu}(p)$.}
\label{fig:photontadpol}
\end{figure}
and as a function of deformation freedom $\kappa$'s ambiguity corrections.
Following the general procedure of dimensional regularization in computing one-loop two-point functions,
we first give the loop-integral
with respect to general integration dimension $D$, then in the following we discuss
the behavior in the specific $D\to 4$ limits.

Out of the above vertices we can read out from  Fig.\ref{fig:photontadpol} the following
loop-integrals
\begin{equation}
\begin{split}
T^{\mu\nu}(p)=&\frac{1}{2}\mu^{d-D}\int\,\frac{d^D \ell}{(2\pi)^D}
\frac{-ig_{\rho\sigma}}{\ell^2}\Gamma^{\mu\nu\rho\sigma}(p,-p,\ell,-\ell)\\
=&e^2\tau^{\mu\nu}\;\mu^{d-D} \int\,\frac{d^D \ell}{(2\pi)^D}\frac{\ell^2}{\ell^2}+e^2{\mathcal T}^{\mu\nu\rho\sigma}\mu^{d-D} \int\,\frac{d^D \ell}{(2\pi)^D}\frac{\ell_\rho \ell_\sigma}{\ell^2}f_{\star_2}^2(p,\ell).
\end{split}
\label{3.3}
\end{equation}
Since the first integral in the above equation (\ref{3.3}) vanishes according to the dimensional regularization prescription~\cite{Leibbrandt:1975dj}, the only remaining integral is the second one. The tensors $\tau^{\mu\nu}$ and ${\mathcal T}^{\mu\nu\rho\sigma}$ are given in Appendix B.

\subsection{Four-photon-tadpole contributions in the limit $D\to 4-\epsilon$}
The single tadpole integral
\begin{equation}
I^{\mu\nu}=\int\,\frac{d^D \ell}{(2\pi)^D}\frac{\ell^\mu\ell^\nu}{\ell^2},
\label{3.3}
\end{equation}
left in \eqref{3.3} boils down to
\begin{equation}
I^{\mu\nu}_{D\to 4}=\frac{1}{6\pi^2}(\theta p)^{-4}\cdot\left(g^{\mu\nu}-4\frac{(\theta p)^\mu(\theta p)^\nu}{(\theta p)^2}\right).
\label{3.13}
\end{equation}
at the $D\to 4$ limit~\cite{Horvat:2015aca}. Combining the partial tensor reduction with the master integral at $D\to 4$ we obtain
\begin{equation}
\begin{split}
T^{\mu\nu}(p)&=\frac{e^2}{3\pi^2}\bigg\{\Big[g^{\mu\nu}p^2-p^{\mu}p^{\nu}\Big]\left(\frac{tr\theta\theta}{(\theta p)^4}+4\frac{(\theta\theta p)^2}{(\theta p)^6}\right)\frac{1}{4}\big(\kappa_4-1\big)
\\&+(\theta p)^\mu (\theta p)^\nu\frac{1}{(\theta p)^4} \,\big(2\kappa^2-4\kappa+6\kappa_1+2\kappa_2-2\kappa_3+\kappa_4-1\big)
\\&+\Big[g^{\mu\nu}(\theta p)^2-(\theta\theta)^{\mu\nu}p^2
+ p^{\{\mu}(\theta\theta p)^{\nu\}}\Big]\frac{1}{(\theta p)^4}\big(2\kappa^2-2\kappa+\kappa_1+\kappa_2\big)
\\&+\Big[(\theta\theta)^{\mu\nu}(\theta p)^2+(\theta\theta p)^\mu(\theta\theta p)^\nu\Big]\frac{2p^2}{(\theta p)^6}\big(\kappa^2-2\kappa+\kappa_1+\kappa_2\big)
\\&+(\theta p)^{\{\mu} (\theta\theta\theta p)^{\nu\}}
\frac{p^2}{(\theta p)^6}\big(2\kappa-\kappa_1-\kappa_2\big)\bigg\},
\end{split}
\label{3.4}
\end{equation}
where we immediately notice the absence of UV and logarithmic divergent terms contrary to the photon-bubble-diagram results \cite{Horvat:2013rga}.


\section{Summing over the bubble and the tadpole diagrams in $D\to 4-\epsilon$}

In this section we present the full quadratic IR divergences in thew 1-loop corrections to the photon polarization tensor by summing up the photon-tadpole contributions \eqref{3.4} and those from the photon bubble diagrams result \eqref{4.2}.

Working out the arithmetics we get the following photon-bubble plus photon-tadpole sum for the photon polarization tensor in the IR regime:
\begin{eqnarray}
\Pi^{\mu\nu}(p)\Big|_{\rm IR}&=&\mathcal B^{\mu\nu}(p)\Big|_{\rm IR} + T^{\mu\nu}(p)\Big|_{\rm IR}
\label{4.5}\\
&=&\frac{e^2}{(4\pi)^2}\bigg\{\Big[g^{\mu\nu}p^2-p^\mu p^\nu\Big]B_{1_{\rm sum}}(p)
+(\theta p)^\mu (\theta p)^\nu B_{2_{\rm sum}}(p)
\nonumber\\&+&\Big[g^{\mu\nu}(\theta p)^2-(\theta\theta)^{\mu\nu}p^2
+ p^{\{\mu}(\theta\theta p)^{\nu\}}\Big]B_{3_{\rm sum}}(p)
\nonumber\\&+&\Big[(\theta\theta)^{\mu\nu}(\theta p)^2+(\theta\theta p)^\mu(\theta\theta p)^\nu\Big]B_{4_{\rm sum}}(p) + (\theta p)^{\{\mu} (\theta\theta\theta p)^{\nu\}} B_{5_{\rm sum}}(p)\bigg\}.
\nonumber
\end{eqnarray}
A summation over the leading UV/IR mixing terms in the bubble and
tadpole diagrams provides results for overall UV/IR mixing (i.e. quadratic IR divergence):
\begin{gather}
\begin{split}
B_{1_{\rm sum}}(p)&\sim-\frac{4}{3}\frac{1}{(\theta p)^4}\Big(tr\theta\theta+4\frac{(\theta\theta p)^2}{(\theta p)^2}\Big)\Big(2(\kappa-1)^2+\kappa_4-1\Big),
\\
B_{2_{\rm sum}}(p)&\sim+\frac{8}{3}\frac{1}{(\theta p)^4}\Big(\kappa^2+2\kappa+12\kappa_1+4\kappa_2-4\kappa_3+2\kappa_4-5\Big)
\\&\hspace{4.5mm}+\frac{16}{3}\frac{p^2}{(\theta p)^6}\Big(tr\theta\theta+6\frac{(\theta\theta p)^2}{(\theta p)^2}\Big)\Big(\kappa-1\Big)^2,
\\
B_{3_{\rm sum}}(p)&\sim-\frac{4}{3}\frac{1}{(\theta p)^4}\Big(9\kappa^2-2\kappa+1-4\kappa_1-4\kappa_2\Big),
\\
B_{4_{\rm sum}}(p)&\sim-\frac{16}{3}\frac{p^2}{(\theta p)^6}\Big(5\kappa^2-2\kappa+1-2\kappa_1-2\kappa_2\Big),
\\
B_{5_{\rm sum}}(p)&\sim+\frac{16}{3}\frac{p^2}{(\theta p)^6}\Big(4\kappa^2-4\kappa+2-\kappa_1-\kappa_2\Big).
\end{split}
\label{4.6}
\end{gather}
It is then straightforward to find that $\kappa=\kappa_4=(\kappa_1+\kappa_2)/2=1$ and $\kappa_3=2\kappa_1+2$ sends all $B_{i_{\rm sum}}$s to zero. Thus the quadratic IR divergence is fully controllable in the sum over one loop photon self-interacting bubble and tadpole diagrams.

\section{Discussion and conclusions}


We present our results on four photon self-interaction in the SW mapped noncommutative U(1) gauge theory and its physical effect through one-loop photon self-interaction tadpole diagram Our results show that the NC massless tadpole integrals are solely quadratically IR divergent, which makes it the potential origin of all quadratic IR divergence as the bubble diagram contains tadpole integrals too.

The successful implementation of a large variety of gauge symmetry inspired freedom parameter $\kappa,\kappa_{i=1...4}$ enables us to control all the quadratic IR divergences using an unique set of these parameters
\begin{equation}
\kappa=\kappa_4=(\kappa_1+\kappa_2)/2=1,\;\;
\kappa_3=2\kappa_1+2.
\label{5.1}
\end{equation}
This choice leaves considerable UV divergences in the bubble diagram on the other hand, which would require an unknown number of gauge-invariant yet nonlocal counter-terms. Thus the authors consider any UV divergence cancellation highly plausible.

Simultaneous elimination of UV plus log and IR divergences can be obtained in a slightly different model \cite{Horvat:2015aca}.

\section{Acknowledgments}
The work by Raul Horvat and Jinagyang You has been fully supported by Croatian Science Foundation under the project (IP-2014-09-9582). The work/project of Josip Trampetic is conducted under the European Commission and the Croatian Ministry of Science, Education and Sports Co-Financing Agreement No. 291823. In particular J.T. acknowledges project financing by The Marie Curie FP7-PEOPLE-2011-COFUND program NEWFELPRO: Grant Agreement No. 69. J.T. would also like to acknowledge Theoretical Physics Divisions at MPI-Munich and at CERN for hospitality.
JY would like to acknowledge Alexander von Humboldt Foundation and COST Action MP1405 for partially supporting his participation of the Corfu Summer Institute 2015 as well as the organizers of the Corfu Summer Institute 2015 for hospitality.
We would like to thank L. Alvarez-Gaume, P. Aschieri, D. Blaschke, M. Dimitrijevi\'c \'Ciri\'c, J. Erdmenger, M. Hanada,  W. Hollik, A.Ilakovac, T. Juri\'c, C. P. Martin, P. Schupp and R. Szabo for fruitful discussions. A great deal of computation was done by using ${\rm Mathematica}$~8.0~\cite{mathematica} plus the tensor algebra package xAct~\cite{xAct}. Special thanks to A. Ilakovac and D. Kekez for the computer software and hardware support.

\appendix

\section{Generalized star products}

The generalized star-products based on the constant Moyal deformation parameter $\theta^{ij}$ bear a relatively common form in momentum space
\begin{equation}
\left[f_1...f_n\right]_{\mathcal M}\left(x\right)=\int\prod\limits_{i=1}^n \frac{d^dp_i}{(2\pi)^d} \prod\limits_{i=1}^n\tilde f_i(p_i)\exp\bigg[-i\Big(\sum\limits_{i=1}^n p_i\Big)x\bigg]{\cal F}\left(p_1,...,p_n;\theta^{ij}\right).
\end{equation}
The examples relevant to this paper are
\begin{equation}
\begin{split}
(f\star g)(x)&=\int\,e^{-i(p+q)x}\tilde f(p)\tilde g(q)f_{\star}\left(p,q\right),\;\;
\\
(f\star_2 g)(x)&=\int\,e^{-i(p+q)x}\tilde f(p)\tilde g(q)f_{\star_2}\left(p,q\right),
\\
[f g h]_{\star_3}(x)&=\int\,e^{-i(p+q+k)x}\tilde f(p)\tilde g(q)\tilde h(k)f_{\star_3}\left(p,q,k\right),\;\;
\\
[f g h]_{\star_{3'}}(x)&=\int\,e^{-i(p+q+k)x}\tilde f(p)\tilde g(q)\tilde h(k)f_{\star_{3'}}\left(p,q,k\right),
\\
[f g h]_{\mathcal M_{\rm (I,II)}}(x)&=\int\,e^{-i(p+q+k)x}\tilde f(p)\tilde g(q)\tilde h(k)f_{\rm (I,II)}\left(p,q,k\right).
\label{2.7}
\end{split}
\end{equation}
with
\begin{gather}
\begin{split}
f_{\star}(p,q)&=\exp\Big(i\frac{p\theta q}{2}\Big),\;\;f_{\star_2}(p,q)=\frac{\sin\frac{p\theta q}{2}}{\frac{p\theta q}{2}},
\\
f_{\star_{3}}\left(p,q,k\right)&=\frac{\sin \frac{p\theta k}{2}\sin(\frac{p\theta q}{2}+\frac{p\theta k}{2})}{(\frac{p\theta q}{2}+\frac{p\theta k}{2})(\frac{p\theta k}{2}+\frac{q\theta k}{2})}+\frac{\sin \frac{q\theta k}{2}\sin(\frac{p\theta q}{2}-\frac{q\theta k}{2})}{(\frac{p\theta q}{2}-\frac{q\theta k}{2})(\frac{p\theta k}{2}+\frac{q\theta k}{2})},
\\
f_{\star_{3'}}\left(p,q,k\right)&=\frac{\cos(\frac{p\theta q}{2}+\frac{p\theta k}{2}-\frac{q\theta k}{2})-1}{(\frac{p\theta q}{2}+\frac{p\theta k}{2}-\frac{q\theta k}{2})\frac{q\theta k}{2}}-\frac{\cos(\frac{p\theta q}{2}+\frac{p\theta k}{2}+\frac{q\theta k}{2})-1}{(\frac{p\theta q}{2}+\frac{p\theta k}{2}+\frac{q\theta k}{2})\frac{q\theta k}{2}},
\end{split}
\label{A8}
\end{gather}
and
\begin{gather}
\begin{split}
f_{\rm (I)}\left(p,q,k\right)&=\frac{2}{(\frac{p\theta q}{2}-\frac{p\theta k}{2}-\frac{q\theta k}{2})(\frac{p\theta q}{2}+\frac{p\theta k}{2}-\frac{q\theta k}{2})(\frac{p\theta q}{2}+\frac{p\theta k}{2}+\frac{q\theta k}{2})}
\\&+\frac{\cos(\frac{p\theta q}{2}-\frac{p\theta k}{2}-\frac{q\theta k}{2})}{2\frac{p\theta q}{2}(\frac{p\theta q}{2}-\frac{q\theta k}{2})(\frac{p\theta q}{2}-\frac{p\theta k}{2}-\frac{q\theta k}{2})}
+\frac{\cos(\frac{p\theta q}{2}+\frac{p\theta k}{2}-\frac{q\theta k}{2})}{2\frac{q\theta k}{2}(\frac{p\theta q}{2}-\frac{q\theta k}{2})(\frac{p\theta q}{2}+\frac{p\theta k}{2}-\frac{q\theta k}{2})}
\\&+\frac{\cos(\frac{p\theta q}{2}+\frac{p\theta k}{2}+\frac{q\theta k}{2})}{2\frac{p\theta q}{2}\frac{q\theta k}{2}(\frac{p\theta q}{2}+\frac{p\theta k}{2}+\frac{q\theta k}{2})},
\\
f_{\rm (II)}\left(p,q,k\right)&=\frac{\sin\left(\frac{p\theta q}{2}-\frac{q\theta k}{2}\right)\sin\frac{p\theta k}{2}}{\frac{p\theta q}{2}\frac{p\theta k}{2}\left(\frac{p\theta q}{2}-\frac{q\theta k}{2}\right)\left(\frac{p\theta k}{2}+\frac{q\theta k}{2}\right)}-\frac{\sin\left(\frac{p\theta q}{2}+\frac{p\theta k}{2}\right)\sin\frac{q\theta k}{2}}{\frac{p\theta q}{2}\frac{q\theta k}{2}\left(\frac{p\theta q}{2}+\frac{p\theta k}{2}\right)\left(\frac{p\theta k}{2}+\frac{q\theta k}{2}\right)}.
\end{split}
\label{A8}
\end{gather}

\section{Tensor reduction and integration results}

The partial $\tau(\mathcal T)$-tensor reduction results are listed bellow
\begin{equation}
\begin{split}
\tau^{\mu\nu}=&-\frac{1}{2D}\Big\{\left[g^{\mu\nu}p^2-p^{\mu}p^{\nu}\right](\tr\theta\theta)(\kappa_3-1)
\\&+\left[g^{\mu\nu}(\theta p)^2-(\theta\theta)^{\mu\nu}p^2
+ p^{\{\mu}(\theta\theta p)^{\nu\}}\right]4(\theta p)^2(\kappa_1-\kappa_2)
\\&+(\theta p)^{\mu}(\theta p)^{\nu}\Big(4+(1-\kappa_3)D(D-1)-16\kappa+8\kappa^2+8(D-1)(\kappa_1-\kappa_2)+4\kappa_4\Big)\Big\},
\end{split}
\label{B12}
\end{equation}
\begin{equation}
\begin{split}
\mathcal T^{\mu\nu\rho\sigma}=&-\frac{1}{2}\Big\{\Big(g^{\mu\nu}(\theta p)^{\rho}(\theta p)^{\sigma}+\theta^{\mu\rho}p^\nu(\theta p)^\sigma+\theta^{\nu\rho}p^\mu(\theta p)^\sigma+\theta^{\mu\rho}\theta^{\nu
\sigma}p^2\Big)
\\&\hspace{6mm}\cdot 2\Big((D-3)\kappa^2-2\kappa+\kappa_1+\kappa_2\Big)
\\&+\Big(2g^{\mu\nu}p^{\rho}(\theta\theta p)^{\sigma}-g^{\mu\rho}p^{\nu}(\theta\theta p)^{\sigma}-g^{\nu\rho}p^{\mu}(\theta\theta p)^{\sigma}
\\&-p^\mu(\theta\theta)^{\nu\rho}p^\sigma-p^\nu(\theta\theta)^{\mu\rho}p^\sigma
+g^{\mu\rho}\theta^{\nu\sigma}p^2+g^{\nu\rho}\theta^{\mu\sigma}p^2\Big)\cdot\big(2\kappa-\kappa_1-\kappa_2\big)
\\&+\Big(g^{\mu\rho}(\theta p)^\nu(\theta p)^\sigma+g^{\nu\rho}(\theta p)^\mu(\theta p)^\sigma+\theta^{\mu\rho}(\theta p)^\nu p^\sigma+\theta^{\nu\rho}(\theta p)^\mu p^\sigma\Big)
\\&\hspace{5mm}\cdot\Big(-1-2\kappa_3+\kappa_4+(2+D)\kappa_1+(D-2)(\kappa_2-2\kappa)+4\kappa^2\Big)
\\&+\Big(g^{\mu\nu}g^{\rho\sigma}(\theta p)^2+(\theta\theta)^{\mu\nu}(p^\rho p^\sigma-p^2 g^{\rho\sigma})+(p^\mu(\theta\theta p)^\nu+p^\nu(\theta\theta p)^\mu)g^{\rho\sigma}
\\&-g^{\mu\rho}g^{\nu\sigma}(\theta p)^2-g^{\mu\rho}(\theta\theta p)^\nu p^\sigma-g^{\nu\rho}(\theta\theta p)^\mu p^\sigma\Big)\cdot 2\kappa^2
\\&+(\theta p)^\mu(\theta p)^\nu g^{\rho\sigma}\cdot 4\Big(\kappa_1+\kappa_2-2\kappa+(D-1)\kappa_4\Big)
\\&-\left[g^{\mu\nu}p^2-p^{\mu}p^{\nu}\right](\theta\theta)^{\rho\sigma}(\kappa_4-1)\Big\}.
\end{split}
\label{B14}
\end{equation}

\section{Model (II)}
\subsection{Interaction}
A second type of Moyal deformed four photon self-coupling can be obtained by inverting the known field strength solution for the inverted SW map~\cite{Mehen:2000vs,Trampetic:2015zma}:
\begin{equation}
\begin{split}
F^{e^3}_{\mu\nu_{\rm (II)}}&(x)_{\kappa,\kappa'_1,\kappa'_2}
=e^3\theta^{ij}\theta^{kl} \Big[\kappa'_1\left(f_{\mu i}\star_2\left(f_{jk}\star_2 f_{l\nu}\right)+f_{l\nu}\star_2\left(f_{jk}\star_2 f_{\mu i}\right)-\left[f_{\mu i}f_{jk}f_{l\nu}\right]_{\star_3}\right)
\\&-\kappa'_2\big((a_i\star_2\partial_j f_{\mu k})\star_2 f_{\nu l}+(a_i\star_2\partial_j f_{\nu l})\star_2 f_{\mu k}-[a_i\partial_j (f_{\mu k}f_{\nu l})]_{\star_3}\big)
\\&- \kappa a_i\star_2\partial_j\left(f_{\mu k}\star_2 f_{\nu l}\right)+(a_i\star_2\partial_j a_k)\star_2\partial_l f_{\mu\nu}
\\&+a_i\star_2(\partial_j a_k\star_2\partial_l f_{\mu\nu})+a_i\star_2(a_k\star_2\partial_j\partial_l f_{\mu\nu})-[a_i\partial_j a_k\partial_l f_{\mu\nu}]_{\star_{3}}
\\&-\frac{1}{2}\Big(a_i\star_2(\partial_k a_j\star_2\partial_l f_{\mu\nu})+(a_i\star_2\partial_k a_j)\star_2\partial_l f_{\mu\nu}-[a_i\partial_k a_j\partial_l f_{\mu\nu}]_{\star_3}
+[a_ia_k\partial_j\partial_l f_{\mu\nu}]_{\star_3}\Big)\Big],
\label{C1}
\end{split}
\end{equation}
which yields the following interaction
\begin{equation}
\begin{split}
S^{e^2}_{\rm (II)}=&-\frac{e^2}{4}\theta^{ij}\theta^{kl}\int\,\kappa^2(f_{\mu i}\star_2 f_{\nu j})(f^\mu_{\;\,\; k}\star_2 f^\nu_{\;\,\; l})-\kappa(f_{ij}\star_2 f_{\mu\nu})(f^\mu_{\;\,\; k}\star_2 f^\nu_{\;\,\; l})
\\&+2\kappa'_1\left(2f_{\mu i}\star_2(f_{jk}\star_2 f_{l\nu})-[f_{\mu i}f_{jk}f_{l\nu}]_{\star_3}\right)
\\&-4\kappa'_2\left((a_i\star_2\partial_j f_{\mu k})\star_2 f_{\nu l}-[a_i\partial_j f_{\mu k}f_{\nu l}]_{\star_3}\right)
\\&-\frac{1}{4}f^{\mu\nu}\left(3f_{ik}\star_2\left(f_{jl}\star_2 f_{\mu\nu}\right)-2\left[f_{ik}f_{jl}f_{\mu\nu}\right]_{\star_3}\right)
\\&+\frac{1}{8}f^{\mu\nu}\left(2f_{ij}\star_2\left(f_{kl}\star_2 f_{\mu\nu}\right)-\left[f_{ij}f_{kl}f_{\mu\nu}\right]_{\star_3}\right)
\\&-\frac{1}{4}\theta^{pq}\theta^{rs}\left[\partial_k f_{ri}\partial_j f_{lp}\partial_q\partial_s f_{\mu\nu}+\partial_i\partial_r f_{jk}\partial_s(f_{lp}\partial_q f_{\mu\nu})\right]_{\mathcal M_{\rm (II)}}.
\end{split}
\label{C2}
\end{equation}
After the same manipulation as in the first model we have
\begin{equation}
\begin{split}
S^{e^2}_{\rm (II)_{\kappa,\kappa'_1,\kappa'_2,\kappa'_3,\kappa'_4}}=&-\frac{e^2}{4}\theta^{ij}\theta^{kl}\int\,\kappa^2(f_{\mu i}\star_2 f_{\nu j})(f^\mu_{\;\,\; k}\star_2 f^\nu_{\;\,\; l})-\kappa(f_{ij}\star_2 f_{\mu\nu})(f^\mu_{\;\,\; k}\star_2 f^\nu_{\;\,\; l})
\\&+2\kappa'_1\left(2f_{\mu i}\star_2(f_{jk}\star_2 f_{l\nu})-[f_{\mu i}f_{jk}f_{l\nu}]_{\star_3}\right)
\\&-4\kappa'_2\left((a_i\star_2\partial_j f_{\mu k})\star_2 f_{\nu l}-[a_i\partial_j f_{\mu k}f_{\nu l}]_{\star_3}\right)
\\&-\frac{\kappa'_3}{4}f^{\mu\nu}\left(3f_{ik}\star_2\left(f_{jl}\star_2 f_{\mu\nu}\right)-2\left[f_{ik}f_{jl}f_{\mu\nu}\right]_{\star_3}\right)
\\&+\frac{\kappa'_4}{8}f^{\mu\nu}\left(2f_{ij}\star_2\left(f_{kl}\star_2 f_{\mu\nu}\right)-\left[f_{ij}f_{kl}f_{\mu\nu}\right]_{\star_3}\right)
\\&-\frac{1}{4}\theta^{pq}\theta^{rs}\left[\partial_k f_{ri}\partial_j f_{lp}\partial_q\partial_s f_{\mu\nu}+\partial_i\partial_r f_{jk}\partial_s(f_{lp}\partial_q f_{\mu\nu})\right]_{\mathcal M_{\rm (II)}}.
\end{split}
\label{C4}
\end{equation}
Note that $\kappa$-terms are identical to the model (I) in the main text, as they should be,
giving in momentum space the following Feynman rule for the model (II):
\begin{equation}
\begin{split}
\Gamma^{\mu_1\mu_2\mu_3\mu_4}_{\rm (II)}\left(p_1,p_2,p_3,p_4\right)=&-i\frac{e^2}{4}\Big(\kappa^2\Gamma_{\rm A}^{\mu_1\mu_2\mu_3\mu_4}\left(p_1,p_2,p_3,p_4\right)
\\&+\kappa\Gamma_{\rm B}^{\mu_1\mu_2\mu_3\mu_4}\left(p_1,p_2,p_3,p_4\right)
+\kappa'_1{\Gamma'}_1^{\mu_1\mu_2\mu_3\mu_4}\left(p_1,p_2,p_3,p_4\right)\\&+\kappa'_2{\Gamma'}_2^{\mu_1\mu_2\mu_3\mu_4}\left(p_1,p_2,p_3,p_4\right)
+\kappa'_3{\Gamma'}_3^{\mu_1\mu_2\mu_3\mu_4}\left(p_1,p_2,p_3,p_4\right)
\\&+\kappa'_4{\Gamma'}_4^{\mu_1\mu_2\mu_3\mu_4}\left(p_1,p_2,p_3,p_4\right)+{\Gamma'}_5^{\mu_1\mu_2\mu_3\mu_4}\left(p_1,p_2,p_3,p_4\right)\Big)
\\&+{\rm all\; S_4\; permutations\; over}\; \{p_i\}{\rm \; and}\; \{\mu_i\}{\rm \; simutaneously}.
\end{split}
\label{C5}
\end{equation}
Details of these vertices are given in~\cite{Horvat:2015aca}.

\subsection{Four-photon-tadpole contributions in the limit $D\to 4-\epsilon$}

The exact tadpole diagram evaluation procedure given in the main text gives following result for the model (II)
\begin{equation}
\begin{split}
T^{\mu\nu}_{\rm (II)}(p)&=\frac{e^2}{3\pi^2}\bigg\{\Big[g^{\mu\nu}p^2-p^{\mu}p^{\nu}\Big]\left(\frac{tr\theta\theta}{(\theta p)^4}+4\frac{(\theta\theta p)^2}{(\theta p)^6}\right)\frac{1}{4}\big(2\kappa'_3+\kappa'_4-3\big)
\\&+(\theta p)^\mu (\theta p)^\nu \frac{2}{(\theta p)^4} \,\big(\kappa^2-2\kappa+2\kappa'_1
+2\kappa'_2-\kappa'_3\big)
\\&+\Big[g^{\mu\nu}(\theta p)^2-(\theta\theta)^{\mu\nu}p^2
+ p^{\{\mu}(\theta\theta p)^{\nu\}}\Big]\frac{1}{(\theta p)^4}\big(2\kappa^2-2\kappa+2\kappa'_2\big)
\\&+\Big[(\theta\theta)^{\mu\nu}(\theta p)^2+(\theta\theta p)^\mu(\theta\theta p)^\nu\Big]\frac{2p^2}{(\theta p)^6}\big(\kappa^2-2\kappa+2\kappa'_2\big)
\\&+(\theta p)^{\{\mu} (\theta\theta\theta p)^{\nu\}} \frac{2p^2}{(\theta p)^6}
\big(\kappa-\kappa'_2\big)\bigg\}.
\end{split}
\label{C7}
\end{equation}

Tensor structure remain exactly the same as for the photon-bubble-diagram in Fig 2 from \cite{Horvat:2013rga}, as one would expect. Due to the absence of UV and logarithmic divergent terms the tadpole contribution from the first model (\ref{3.4}) can be made equal to that of the model (II) when setting
\begin{eqnarray}
\kappa_1+\kappa_2&=&2\kappa'_2,
\nonumber\\
4\kappa_1-2\kappa_3&=&4\kappa'_1-4\kappa'_3-\kappa'_4+3,
\nonumber\\
\kappa_4&=&2\kappa'_3+\kappa'_4-2,
\label{C8}
\end{eqnarray}
and in particular $T^{\mu\nu}(p)=T^{\mu\nu}_{\rm (II)}(p)$, for $\kappa_i=\kappa'_i=1, \forall i$.

Summing up (\ref{4.2}) and (\ref{C7}) we obtain the Model (II) coefficients:
\begin{gather}
\begin{split}
B_{1_{\rm sum\,(II)}}(p)&\sim-\frac{4}{3}\frac{1}{(\theta p)^4}\Big(tr\theta\theta+4\frac{(\theta\theta p)^2}{(\theta p)^2}\Big)\Big(2(\kappa-1)^2-2\kappa'_3-\kappa'_4+3\Big),
\\
B_{2_{\rm sum\,(II)}}(p)&\sim+\frac{8}{3}\frac{1}{(\theta p)^4}\Big(\kappa^2+2\kappa-3+8\kappa'_1+8\kappa'_2-4\kappa'_3\Big)
\\&\hspace{4.5mm}+\frac{16}{3}\frac{p^2}{(\theta p)^6}\Big(tr\theta\theta+6\frac{(\theta\theta p)^2}{(\theta p)^2}\Big)\Big(\kappa-1\Big)^2,
\\
B_{3_{\rm sum\,(II)}}(p)&\sim-\frac{4}{3}\frac{1}{(\theta p)^4}\Big(9\kappa^2-2\kappa+1-8\kappa'_2\Big),
\\
B_{4_{\rm sum\,(II)}}(p)&\sim-\frac{16}{3}\frac{p^2}{(\theta p)^6}\Big(5\kappa^2-2\kappa+1-4\kappa'_2\Big),
\\
B_{5_{\rm sum\,(II)}}(p)&\sim+\frac{16}{3}\frac{p^2}{(\theta p)^6}\Big(4\kappa^2-4\kappa+2-2\kappa'_2\Big).
\end{split}
\label{C10}
\end{gather}
Note that IR divergence in the first model of the coefficients $B_{3,4,5}$ from (\ref{4.6}) depends
on $\kappa_1+\kappa_2$, while here only on $\kappa'_2$. Like in the first model (\ref{5.1}), model (II) also has a choice of freedom parameters
\begin{equation}
\kappa=\kappa'_2=1,\;\;\kappa'_3=2\kappa'_1+2,\;\;\kappa'_4=3-2\kappa'_3,
\end{equation}
which ensures full quadratic IR divergence cancellation for any $\theta^{ij}$.

\end{document}